\UseRawInputEncoding
\documentclass[runningheads]{llncs}

\usepackage[english]{babel}

\usepackage[tbtags]{amsmath}
\usepackage{latexsym, amssymb, amsmath, amsfonts, array, amscd, hhline, mathrsfs, bussproofs}

\usepackage{etex}

\usepackage[all,2cell]{xy}

\usepackage{makeidx,fancybox,tikz}
\usetikzlibrary{automata, positioning, arrows,patterns}
\usepackage{graphicx}

\usepackage{xcolor}

\usepackage{alltt}
\usepackage{float}

\usepackage{url}
\usepackage{doi}

\newcommand{\ednote}{\triangle}

\newcommand{\contr}{{$\stackrel{c}{\rightarrow}$}~}

\newcommand{\empt}{$\epsilon$}
\newcommand{\concat}{\mbox{\ \vspace{5ex}\\{\tt {+\hspace{-0.5ex}+}}}}

\newcommand{\zb}{\hspace{-0.5mm}}
\newcommand{\zbb}{\hspace{-0.25mm}}

\newcommand{\dddz}{{{\texttt {\scriptsize{\#}}}\zbb}}

\def\quote{\raisebox{-0.4ex}{\rotatebox[origin=c]{13}{${}^{{\tiny{\prime}}}$}\hspace{-0.3ex}}}
\def\bquote{\hspace{-0.2ex}{\rotatebox[origin=c]{13}{${}^{{\tiny{\prime}}}$}}}

\newcommand{\Nat}{\mathbb{N}}

\newcommand{\programenvironment}{\programmode%
	\def\par{\leavevmode\endgraf}\obeylines\nobreak%
	\programmode}

\newcommand{\programmode}{\normalsize\tt
	\catcode`\_=12 \catcode`\?=12 \catcode`\.=12 \catcode`\,=12
	\catcode`\;=12 \catcode`\:=12 \catcode`\@=12 \catcode`\~=12
        \catcode`\#=12 \catcode`\&=12      
	\obeyspaces\frenchspacing}%

\newenvironment{programintext}{\programenvironment}{}

\newenvironment{program}{\setlength{\partopsep}{0mm}\setlength{\topsep}{0mm}
	\begin{trivlist}\item[]
\begin{minipage}{0.90\textwidth}
	\vspace{1mm}
	\begin{programintext}
	}{\end{programintext}
	\vspace{1mm}
	\end{minipage}
	\end{trivlist}
	\noindent}

{\catcode`\^^I=\active \obeyspaces\global\def^^I{ }}
{\obeyspaces\global\let =\ } 
{\catcode`\`=\active \gdef`{\relax\lq}}

\usepackage{color}            

\newcommand{\ie}{{\it i.e.},~}

\newcommand{\st}{{\it s.t.}~}

\newcommand{\wrt}{{\it w.r.t.}~}
\newcommand{\ddi}{i}

\newcommand{\sectionname}{Sec.}

\newcommand{\vara}{$\mathtt{s_{a}}$}
\newcommand{\para}{\mathtt{{_{\dddz}}s_{a}}}
\newcommand{\varb}{$\mathtt{s_{b}}$}
\newcommand{\parb}{\mathtt{{_{\dddz}}s_{b}}}
\newcommand{\varx}{$\mathtt{p}$}
\newcommand{\vary}{$\mathtt{y}$}
\newcommand{\varz}{$\mathtt{z}$}
\newcommand{\varq}{$\mathtt{q}$}
\newcommand{\parystr}{\mathtt{{_{\dddz}}y_{str}}}
\newcommand{\parpatt}{\mathtt{{_{\dddz}}x_{\pi}}}
\newcommand{\parystrzero}{\mathtt{{_{\dddz}}y_{str_0}}}
\newcommand{\parxstr}{\mathtt{{_{\dddz}}x_{str}}}
\newcommand{\cons}{{\tt \hspace{0.2ex}:\hspace{0.2ex}}}
\newcommand{\comm}{-{\hspace{0.2ex}}-}
\newcommand{\Nil}{Nil}
\newcommand{\True}{$\mathbf{T}$}
\newcommand{\False}{$\mathbf{F}$}
\newcommand{\Patt}{\mathtt{x_{\pi_0}}}
\newcommand{\ProgNP}{{\tt P}}
\newcommand{\Go}{\tt S}
\newcommand{\scp}{\mathtt{Spec}}
\newcommand{\Code}{\underline}
\newcommand{\SpecTask}{\mathfrak{T}}

\def\frameit{\smallskip \advance \linewidth by -7.5pt \setbox0=\vbox \bgroup
\strut \ignorespaces }

\tikzset{
>=stealth',        
node distance=2cm, 
every state/.style={thick, fill=gray!5}, 
initial text=$ $,                        
}

\begin{document}

\title{On Specialization of a Program Model \newline of Naive Pattern Matching in Strings \newline
{\small{(Extended Abstract)}}
}

\titlerunning{On Specialization of Na{\ddi}ve Pattern Matching in Strings}

\author{Andrei P. Nemytykh\thanks{
The author was partially supported by Russian Academy of Sciences, research project No. AAAA-A19-119020690043-9.
}
\institute{{Program Systems Institute of Russian Academy of Sciences} \\
\email{{nemytykh@math.botik.ru}}
}}

\authorrunning{Andrei P. Nemytykh}

\date{}
\maketitle

\begin{abstract}
We have proved that for any pattern $p$ the tail recursive program model of 
na{\ddi}ve pattern matching may be automatically specialized \wrt the pattern $p$ to a specialized version of the so-called  KMP-algorithm, using the Higman-Kruskal relation that controls the unfolding/folding. Given an input string, the corresponding residual program finds the first occurrence of $p$ in the string in linear time on the string length. The current state of the automated program specialization art based on unfolding/folding is too weak in order to be able to reproduce the proof, done by hands, of the uniform property above, while it known before that program specialization is sometimes able to produce the KMP-algorithm for a few concrete static patterns.

\keywords{Program specialization $\cdot$ Supercompilation $\cdot$ Optimization $\cdot$ KMP-algorithm $\cdot$ Program verification}
\end{abstract}

Proving uniform properties of program optimizers (or transformers) for various computational models is of fundamental value to our understanding of both compilation and computation. Here a property of a given optimizer is said to be \emph{uniform} iff there are input {\it static} arguments of the program to be optimized \st the property holds \emph{for any} of the arguments' \emph{static} values,
while other input arguments may be dynamic. Thus a uniform task is posed \wrt a subset of input arguments when the task is supposed to be solved by an human interested in the uniform property of a program specializer. When one wants to pose the corresponding uniform task above to a program specializer rather than an human then the mentioned static arguments should be redeclared as dynamic ones. This paper concerns itself with solving a uniform task posed to an human since the modern program specializers are unable to solve the task.
The author believes that we are still very far from proving non-trivial uniform properties of optimizers for realistic models of computation. 

 In this extended abstract, we report on a study of some uniform properties of a program specialization method known as Turchin's supercompilation \cite{Tur:86,Turchin:Obninsk90,Sorensen:94,Nem_Pin_Tur,NT:00,Nemytykh:SCP4book,ANepeivoda:MSCP-A}. Namely, the properties of a supercompiler when it specializes a program model of 
 na{\ddi}ve pattern matching in strings \wrt the pattern.
 
 One can apparently deem that any program analyzing character strings uses a predicate testing equality of such two strings as well as a function looking for the first occurrence of a given substring in an input string. Automated exploration of diverse program models of these two functions is an interesting, difficult, and practically important task.
 
 The idea of studying program specialization methods by transforming the programs modeling 
 na{\ddi}ve pattern matching originates 
 from Yo.~Futamura and K.~Nogi (\cite{Futamura:1988}, 1987). Here, by {\it{``na{\ddi}vity''}} of an algorithm is meant its natural essence not messed up by any thought, \ie 
 not incorporating some ingenuity.
 The authors experimented with the program model written in terms of the LISP language, which can be encoded, up to 
 a morphism,  
in 
term rewriting systems 
\cite{TRWs:Barendsen}
based on top-down pattern matching as follows. 
 
 \begin{program}\label{prog:Search}
{\small{
S \{  \comm Search \--- Pattern matching
   \vara\cons\varx,\vara\cons\vary = L(\vara\cons\varx,\vara\cons\vary,\vara\cons\varx,\vary);
   \vara\cons\varx,\varb\cons\vary = S(\vara\cons\varx,\vary);
      \varx,\;\Nil = \False;  \}

L \{  \comm Look for the first pattern symbol \vara inside the string \vary\footnote{The \varx,\,\varq,\,$\mathtt{x}$,\,\vary,\,\varz are variables ranging over the lists and may be indexed. The indexed $\mathtt{s}$-variables range over the symbols. The identifiers \True and \False stand for the logical constants True and False, respectively. The rewriting rules defining functions, here{\tt S} and{\tt L}, are enclosed in curly brackets.\label{param:def}}.
   \vara\cons\varx,\vara\cons\vary,\varq,\varz = L(\varx,\vary,\varq,\varz);
   \vara\cons\varx,\varb\cons\vary,\varq,\varz = S(\varq,\varz);
   \vara\cons\varx, \Nil,\varq,\varz = S(\varq,\varz);
    \Nil,   \vary,\varq,\varz = \True;  \}
}} 
 \end{program}

The worst-case time complexity of this program model is ${\cal O}(|{\tt p}| \times |{\tt y}|)$, where {\tt p} and {\tt y} are input pattern and string, respectively.

The specialization task of our interest is defined as follows. 
 $$\SpecTask(\ProgNP,\Patt)\triangleq \scp(\,\Code{\ProgNP}, \Code{\Go(\Patt},\parystr\Code{)}\,),$$
where $\scp$ is a program specializer, $\ProgNP$ stands for the program above, $\Go$ is its entry function. 
We use the underlining sign to show encoded structures of the program to be specialized. 
This initial configuration takes a {\it static} pattern and {\it dynamic}\footnote{In the sequel the term {\it parameter} is used for dynamic variables, in order to stress that the parameter value is given but unknown 
to the transformer, 
while the variables are to be assigned. The parameters will be prefixed with the sign $\mathtt{{_{\dddz}}}$.} string. 
A pointer moves from left to right along the string $\parystr$, looking for the first occurrence of $\para$ being the first letter of the input pattern $\Patt$. When such an occurrence is found the unscanned segment of $\parystr$ and the pattern are saved as a backtracking point (the first rule of ${\tt S}$), in order to continue the pattern matching in $\parystr$ if a prefix of $\para\,\parystr$ does not coincide with $\Patt$ (see the second rule of {\tt L}). The corresponding model is a tail recursive program. The predicate $\Go$ moves the pointer along the input string $\parystr$. The predicate ${\tt L}$ compares a prefix of $\para\,\parystr$ with the pattern given in its first argument.

\section{The Matiyasevich \& Knuth-Morris-Pratt Algorithm}\label{Sec:M-KMP}

The residual program of the specialization task $\SpecTask(\ProgNP,{\tt \quote{aaab}\quote})$ 
reported by Futamura and Nogi is not only a version of 
na{\ddi}ve pattern matching specialized \wrt the pattern,
but a specialized version of a searching algorithm solving the same task that was apparently discovered independly by Yu.~V. Matiyasevich (1969, 1971 \cite{Matiyasevich:M-KMP}, 1973 \cite{Matiyasevich:M-KMP}), and J.\,H.~Morris and V.\,R.~Pratt (1970, \cite{Morris-Pratt:KMP-70}) \--- published in 1977 by D.~Knuth, J.~Morris and V.~Pratt \cite{Morris-Pratt:KMP-77}.
See also \cite{Shen:ProgTheorems}.

\begin{figure}[h]
\begin{tikzpicture}[scale=0.9]
\draw [fill=gray!20] (2.0,0) rectangle (3.0,0.5);
\draw (3,0) rectangle (3.4,0.5);
\draw (3,0) rectangle (4.5,0.5);
\draw [pattern=north west lines,  pattern color=gray](5,0) rectangle (7,0.5);
\draw [fill=gray!20](7.0,0) rectangle (8.0,0.5);
\draw [fill=gray!50](8.0,0) rectangle (8.4,0.5);
\draw (8.4,0) rectangle (12.7,0.5);
\draw [thick, ->] (7,1.2) -- (7,0.5);
\node at (0.5,0.22) {};
\node at (2.5,0.22) {$q$};
\node at (3.2,0.29) {$d$};
\node at (7.5,0.22) {$q$};
\node at (8.2,0.27) {$c$};
\draw [dashed,thick](4.76,-0.4) -- (4.76,0.9);
\draw (2,-0.3) arc (250:290:3.7cm);
\draw (5,-0.3) arc (260:280:22.2cm);
\node at (3.22,-0.9) {\small {The pattern $p$}};
\node at (8.8,-0.9) {\small {The string $\parystr$}};
\node at (8,1.8) {\small \begin{tabular}{l}The scanned part of the string is to the left of the pointer,\\ the unscanned one is to the right. $c \neq d$.\end{tabular}};
\end{tikzpicture}
\caption{The needed pattern $p$ and the input string $\parystr$.}
\label{fig:M-KMP-algorithm}
\end{figure}
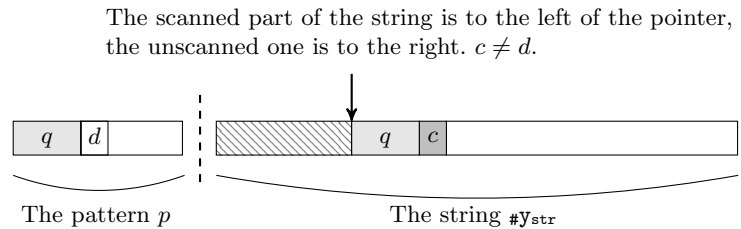

\paragraph{The algorithm M-KMP} in linear time on the pattern length $|p|$ generates firstly a function $f(q)$. 
Let a state defined by the scanned prefix $q$ of $p$ that we are looking for be given. Let $i$ stand for the pointer place of the observed occurrence of a letter unequal to the letter $d$ indexed with $|q|+1$ in $p$.
See Fig.~\ref{fig:M-KMP-algorithm}.
The function allows us, using the only program step, to move the pointer from the $i$-th index to the index $j(i) = i-f(q)$ along the string $\parystr$, before which the pattern $p$ 
cannot have an occurrence.
The function $f(q)$ is discovered from the structure of $p$. Then the direct search starts using $f(q)$.
 
  The worst-case time complexity of the entire algorithm M-KMP is ${\cal O}(|p| + |\parystr|)$.

We name the $f$-function property described in 
the previous 
paragraph the primary property of this function.
Let $l(q)$ be both a prefix and a suffix of the word $q$ then $f(\texttt{\empt}) = 0$,
and if $q \neq$\;\empt: $f(q) = \max\bigl\{ |l(q)| \bigm| l(q) \neq q\bigl\}$, where \empt~is the empty word. 
For example, for $p$
\begin{program}
	   \quote{aab}\quote: $(f(\texttt{\empt})\,=\,0,\,j(i)\,=\,i),\,(f($\quote{a}\quote$)\,=\,0,\,j(i)\,=\,i),$
	                           \;$(f($\quote{aa}\quote$)\,=\,1,\,j(i)\,=\,i-1)$; 
	 \quote{ababa}\quote: $(f(${\empt}$)\,=\,0,\,j(i)\,=\,i),\,(f($\quote{a}\quote$)\,=\,0,\,j(i)\,=\,i),$\\
\vspace{-5.6ex}
	         $(f($\quote{ab}\quote$)\,=\,0,\,j(i)\,=\,i),\,(f($\quote{aba}\quote$)\,=\,1,\,j(i)\,=\,i-1),\,$\\ 
\vspace{-5.6ex}
                              \hspace{-0.7ex}$(f($\quote{abab}\quote$)\,=\,2,\,j(i)\,=\,i-2)$.
\end{program}

 An invariant of the algorithm second stage is that the scanned prefix of $p$ and the unscanned prefix of $\parystr$ coincide. If the unscanned prefix of $\parystr$ coincides with $p$ then we have found the first occurrence of $p$ in $\parystr$. If the string has been finished not satisfying this property then it does not contain the input pattern as a substring. 
 The multiple and overlapping occurrences of concrete substrings in the needed pattern cause the main difficulty in discovering the function $f(q)$.
 
 \paragraph{Short history of the task considered:}\label{history}
 The first Futamura-Nogi results above were obtained in 1987 by means of the generalized partial computation method \cite{Futamura:1988} and published in 1988. Their results are substantially based on 
 exploiting
 negative information\footnote{I.\,e., on an information described using the negation logical connective.} about the parameterized program configurations used. The first paper reporting on successful partial evaluation experiments of specialization of the corresponding models appeared in 1989 \cite{Consel-Danvy:KMP-89}, but in order to obtain desirable results of specializing the models \wrt three pattern samples including the pattern {\tt \quote{abcabcacab}\quote} the paper authors were forced to disrupt the natural essence of 
 na{\ddi}ve pattern matching. Later the task of interest was popular enough, see for examples 
 manual elaborating
 the idea above 
 in the domain of finite trees \cite{Smith:1991},
 a report \cite{PettorossiProietti:KMP96}  on experiments done by means of partial deduction. 
Same technique is used in \cite{PettProRen:DerivKMP05} in order to improve a number of examples of na{\ddi}ve, nondeterministic program models to specialized versions of the M-KMP algorithm, among many others papers presenting a few specialization tasks resulting in similar residual programs.
We have to point out to an interesting work \cite{Bird89:KMP} that calculates, by means of hands, an {\it optimized} version of a general na{\ddi}ve matcher where the input pattern is {\it dynamic}, which results in the general M-KMP algorithm. That approach is based on an algebraic technique and uses not automated tricks including higher order relations decreasing the worst-case time complexity.

 To our knowledge, excluding exhaustive computation for a number of concrete patterns, there was no attempt to describe an analysis of causes of unfold/fold specializing the na{\ddi}ve pattern matching \wrt an {\it arbitrary} concrete pattern $\pi_0$, \ie a {\it static} one, leading to generating the specialized version of the M-KMP algorithm, looking for $\pi_0$. Thus this report is the first one presenting results of such an analysis. Proving this uniform property of program specialization was remained a challenging task for a long time.
 
 \paragraph{Model and Result:} 
 Throughout this paper we assume that the program model of 
 na{\ddi}ve pattern matching is fixed. It is the program {\tt P} given in the introduction above. It is clear that the property of interest depends on \emph{both} the \emph{program model} and the program specializer. Thus this paper addresses a relation between the model and the specializer considered. We claim that
 for \emph{any} given pattern $\pi$ the supercompiler SCP4 \cite{Nemytykh:SCP4book,NT:00} solving the task $\SpecTask(\ProgNP,\pi)$ above generates a residual program being a specialized version of the M-KMP algorithm and running in linear time on the input string length $|\parystr|$ \emph{unknown} to the specializer. The string $\parystr$ is dynamic.
 So we have infinitely many specialization tasks, each of them corresponds to its own $\pi$. We have manually proved the relation above uniformly\footnote{I.e., this single proof holds for each of the tasks, for any fixed pattern. This uniform proof (as well as property) concept is widely used in mathematics.} over these tasks, \ie actually over the pattern set.  
 Given a program specializer $\scp$, in order to automatically prove the relation above one should launch the specializer infinitely many times.
 
 There is another specialization task uniting the tasks above. It is as follows
$$\scp(\,\Code{\ProgNP}, \Code{\Go(}\parpatt, \parystr\Code{)}\,),$$
where both $\parpatt$ and $\parystr$ stand for parameters ranging over {\it dynamic} patterns and {\it dynamic} strings, respectively.
The current state of the automated program specialization art based on unfolding/folding is too weak in order to result in a specialized version of the M-KMP algorithm, when solving this 
task.

\section{Preliminaries}\label{sec:Prelim}

We assume the reader to be familiar with the basics in program specialization.
Let $\cal{A}$ be a finite alphabet of letters, constants. 
Henceforth, $\pi$ stands for a constant pattern unknown to the reader.  See the footnote below.
$\xi_i \notin \cal{A}$ stands for the $i$-th unspecified letter 
of $\pi, 0 < i \leqslant |\pi|$. Here and below for any $\alpha \in A^{*}$\ $|\alpha|$ is the length of the word \--- the number of letters in $\alpha$.
$\pi  = \xi_1 \ldots \xi_i \ldots \xi_{|\pi|}$. 
$\pi_i$ is the $i$-th 
 nonempty
suffix of $\pi$ defined by $\pi_0 = \pi$, for $0 < i < |\pi|$,
$0 < |\pi_i| = |\pi|-i < |\pi|$, $\omega_i$ is the $i$-th prefix of the word $\pi_1$ \st $|\omega_i| =i-1$.
So the following equalities hold for any $i$ \st 
$0 < i < |\pi|$: 
$\pi_1 =  \omega_i\, \pi_i$, $\pi_{(i-1)} = \xi_i\,  \pi_i$.
For example, $\omega_1 =$\;\empt$, \pi_{(|\pi|-1)} = \xi_{|\pi|}$.
In this notation the pattern first letter $\xi_1$ is an exclusive one, since it is appropriately treated by the program {\tt P}. From the point of view of the following reasonings, 
the symbols $\xi_i, \pi_j, \omega_k$ 
are meta-parameters\footnote{One may consider the symbols $\xi_i, \pi_j, \omega_k$ as aliases for the pattern letters and segments.
Note the value of $\xi$ is in ${\cal A}$, while the letter $\xi$ itself is outside of ${\cal A}$.} meaning that their values are unknown to the reader, but 
 fixed 
input data \cite{Turchin:JFP93,Tur_Nem:1995,Nem_Pin_Tur}; while their values are known to the specializer, 
where $\xi_i$ ranges over ${\cal A}$, 
$\pi_j, \omega_k$ range over  ${\cal A}^{*}$.
Sometimes we will briefly call them either letters or words respectively, by default, assuming that their values are unknown to us.
A parameterized expression and a word are defined as follows. The {\it ordinary} {\tt parameter}s, \ie without the meta prefix, were introduced above 
(see Page~\pageref{prog:Search}).
\begin{program} 
 pexpr ::= \Nil | \quote$\nu$\quote\,:\,pexpr | parameter\,:\,pexpr | F(args)
    args ::= pexpr | pexpr, args
 word  ::= \Nil | \quote$\nu$\quote\,:\,word\footnote{Where $\nu \in {\cal A}$, {\tt F} is a function name. We use a widely known abbreviation for the words, for example, \zb\zb\zb{\tt \quote{abcda}\quote} stands for \zb\zb\zb{\tt \quote{a}\quote:\quote{b}\quote:\quote{c}\quote:\quote{d}\quote:\quote{a}\quote:\Nil}. 
 }   
\end{program} 

A configuration, \ie a parameterized expression, containing a function call is said to be active, otherwise it is called passive.
\begin{definition}
 Let a program {\tt P} and its parameterized entry configuration 
 {\tt F(pexpr)} 
 be given. Let ${T_n}$ be a sequence of rooted directed trees\footnote{with edges going from the root.} that are defined recursively.

$T_0 \triangleq 
{\tt F(pexpr)}$. 
Given a tree $T_n$, then $T_{(n+1)}$ is the replacement result of every leaf of $T_n$, labeled with an active configuration $[C]$, with the tree generated by 
the one-step unfolding of $[C]$. 
The sequence $\{T_k\}_{k\ge0}$ is said to be 
the 
    complete
    unfolding tree of the pair  
$\langle {\tt P}, 
{\tt F(pexpr)} 
\rangle$ and denoted with 
${\hat{\tt P}}_{\tt F(pexpr)}$. 
\hfill$\ednote$
\end{definition}

${\hat{\tt P}}_{\tt F(pexpr)}$ 
can be finite or infinite.
${\hat{\tt P}}_{\tt F(pexpr)}\triangleq\{T_k\}_{k\ge0}$ 
is said to be finite iff there exists $k$ \st $T_{(k+1)} = T_k$. 
For any $k \in \Nat$, $T_k$ is a partial computation tree. The sequence ${\hat{\tt P}}_{\tt F(pexpr)}$ can be informally seen as  
$\lim_{i \to \infty} T_i$. I.e., it can be considered as the infinite parallel unfolding of the pair 
$\langle {\tt P}, {\tt F(pexpr)} \rangle$. 
We omit the index of this tree if it is clear from the context of use. For example, all the following definitions make sense for any program {\tt P} and any its entry configurations, so ${\hat{\tt P}}_{\tt F(pexpr)}$ is shortened to ${\hat{\tt P}}$.
We abuse notation and denote this ``limit'' by ${\hat{\tt P}}$. So ${\hat{\tt P}}$ also stands for a finite or infinite tree.
It should be clear from context which definition is intended.

We call a node transient, if the one-step unfolding of the configuration labelling it produces 
the only edge outcoming from the node.
Unless specified otherwise, we assume that all transient nodes are removed from 
the complete
unfolding tree.\footnote{Whenever a transient node is removed then its incoming and outcoming edges are replaced with a single edge labeled with the composition of the predicates labeling the removed edges.} 

Given a tree generated by the one-step unfolding of a configuration, a node in this tree is said to be a pivot node if it is the first node along a path starting at the tree root, and having at least two outcoming edges.

\typeout{
\textbf{? Maybe the remark above on transient node should be commented ...} 
\textbf{Actually, the transient configuration should be removed. See Russian version of this paper.} 
}

Paths generated by a single unfolding step are ordered. This order respects the order of the steps done by the machine meta-interpreting 
{\tt P} and constructing the paths. The paths in ${\hat{\tt P}}$ are lexicographically ordered \wrt the following pairs: the name of the function being specialized; the order numbers of the rewriting rules corresponding to the current unfolding operation along the path considered. Henceforth, we use the order path terminology in ${\hat{\tt P}}$ corresponding to the lexicographical order, unless specified otherwise. 

Let $Q(t_i, t_j)$ stand for a formula of the form $t_i \neq t_j$, where every argument is either an {\tt s}-parameter or a symbol. Here we are interested in the predicates $R(t_1,\ldots,t_n)$ being conjunctions of such elementary inequalities $Q(t_i,t_j)$. Such a predicate  restricts domains of the parameters from its arguments.

Now we extend the configuration concept. See also \cite{Turchin:Report20,Futamura:1988,Turchin:Obninsk90,Futamura:1991}.

\begin{definition}
A parameterized configuration is a pair of the form $\langle {\tt pexpr}, R(p_1,\ldots,p_n) \rangle$, where $p_i$ are {\tt s}-parameters. $R$ is a predicate specifying  {\it ``negative information''} restricting the domains 
of parameters $p_1, \ldots, p_n$ from {\tt pexpr}. 
\end{definition}

\begin{definition}
Let 
     a complete 
unfolding tree ${\hat{\tt P}}$ and parameterized configurations
$[C_1] \triangleq \langle {\tt pexpr}_1, R_1(p_1, \ldots, p_n) \rangle$, 
$[C_2] \triangleq \langle {\tt pexpr}_2, R_2(q_1, \ldots, q_k) \rangle$ labeling nodes in 
${\hat{\tt P}}$ 
be given. 
We say $[C_1]$ covers $[C_2]$ if there is a renaming $\sigma$ of the $[C_1]$ parameters \st  
$\sigma({\tt pexpr}_1) = {\tt pexpr}_2$ and the predicate $R_2(q_1, \ldots, q_k) \Rightarrow R_1(\sigma(p_1), \ldots, \sigma(p_n))$ is identically true.
\end{definition}

\begin{definition}
Let 
     a complete 
unfolding tree ${\hat{\tt P}}$, a path $\mathfrak{t}$ starting at the ${\hat{\tt P}}$ root, and parameterized configurations $[C_1]$, $[C_2]$ along the path $\mathfrak{t}$ be given. We say a segment $\Delta_1$ of $\mathfrak{t}$ covers 
 $[C_1]$ if there is a configuration $[C_0]$ along $\Delta_1$ that covers $[C_1]$. 

Let $T$ be a subtree of ${\hat{\tt P}}$, rooted in $[C_2]$. We say a segment $\Delta_2$ of the path $\mathfrak{t}$, consisting of $[C_2]$-ancestors, covers the subtree $T$ if along any infinite path $\mathfrak{r}_i$ starting at $[C_2]$ there is a configuration $[K_i]$ covered by $\Delta_2$.
\end{definition}

 The following lemmata relating to the na{\ddi}ve pattern matching model given in the introduction are a part of our contribution.
We have proved the statements below and Theorem~\ref{th:on-covering} based on them, assuming, by default, that the unfolding/folding process is managed by the Higman-Kruskal relation \cite{Higman,Kruskal} and other conditions (if given below). 

\begin{lemma}\label{lemma1}
For any $\pi \in {\cal A}^{+}$ the first ${\hat{\tt P}_{{\tt S(}\pi,{\tt \parystr)}}}$-path starting at the root ends at a leaf labeled with the passive configuration {\tt \True} and all pivot configurations along this path generated by the supercompiler SCP4 \cite{NT:00,Nemytykh:SCP4book} form the following finite sequence:
$$[S], [L_1], \ldots , [L_{(n-1)}],$$ 
where $n = |\pi|$, $[S]$ is the initial configuration, $[L_i] \triangleq L(\pi_i, \parystr, \pi, \omega_i\zb\zb\concat \parystr)$\footnote{The sign \zb\zb\concat\ stands for the associative concatenation.}.
The transient configuration $[L_n]$ follows this sequence of the active configurations along the first path; 
and for all $i, j \in \Nat$ \st $0 \le i < j < n$ the inequality $|\pi_i| > |\pi_j|$ holds. \hfill$\ednote$
\end{lemma}

\begin{lemma}\label{lemma2}
For any $\pi \in {\cal A}^{+}$ and for any infinite path $\mathfrak{r}$ 
that starts  
at the root of ${\hat{\tt P}_{{\tt S(}\pi,{\tt \parystr)}}}$ and 
goes
through at least one configuration with a call of the function {\tt L}\footnote{A path not including such a configuration corresponds to the input strings not containing the pattern's first letter.
}
 there is a configuration $[S{\tt\bquote}]$ of the form 
$\langle S(\pi, {\tt \para \cons \parystr)}, R({\tt \para}) \rangle$ \st the root of ${\hat{\tt P}_{{\tt S(}\pi,{\tt \parystr)}}}$ is the only ancestor of $[S{\tt\bquote}]$ with an {\tt S}-call and it is a pivot. 
(See Fig.~\ref{fig3:Prelim}.) 
\end{lemma}

{
\begin{figure}[h]
\[
\def\objectstyle{\scriptstyle} 
\def\labelstyle{\scriptstyle}
\xymatrix @+2mm @R-25pt @C-0pt {
  &
  *[o][F]{\mathbf{\bullet}} 
        \ar[r]^{} 
  & *+[F-:<3pt>]{\texttt{[S]}}
     \ar@{-->}[rr]^{\mathfrak{r}}  
     &
     &    *+[F-:<3pt>]{\texttt{[L]}}
     \ar@{.>}[rrr]^{}  
     \ar[dr]^(.6){\mathfrak{r}} 
     &
     &   
     &
\\
  &
     &  
     &          
     & 
     &  \bullet \ar@{.}@/_0.2pc/[rr]
                \ar@{.>}[dr]^(.7){\mathfrak{r}} 
                \ar@{.}[ddddrr]
     &  
     &  \ar@/^1.5pc/@{.}[dddd] \ar@/^0.5pc/@{.}[dd]
\\ 
  &
     &  
     &          
     & 
     & 
     &  *+[F-:<3pt>]{\texttt{\hspace{-0.3ex}[{\it S}\quote]\hspace{-0.5ex}}} 
                                   \ar@{.}@/_0.3pc/[ur]^(0.9){\texttt{\ \ }}
                                   \ar@{.>}[r]^(.8){\mathfrak{r}}
                                   \ar@{.}@/^0.3pc/[dr]
     &   
\\ 
  &
     &  
     &          
     & 
     &  
     &  \ar@/_0.7pc/@{.}[uu] 
     &
\\ 
  &
     & & & & & &  
\\ 
  &
     & & & & & &  
\\ 
}
\]
%
\vspace{-20pt}
\caption{The complete unfolding tree ${\hat{\tt P}_{{\tt S(}\pi,\ {\tt \parystr)}}}$. (Here and below the dashed and dotted arrows indicate  segments of the paths, which may contain a number of edges, while each continuous arrow indicates the only edge.)}\label{fig3:Prelim}
\end{figure}
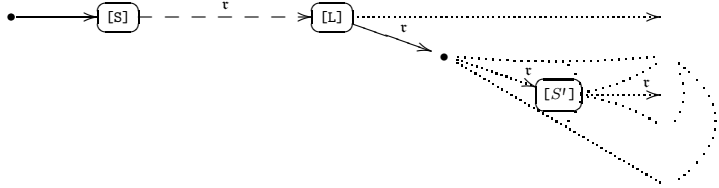
}

\begin{lemma}\label{lemma3}
For any $\pi \in {\cal A}^{+}$ and for any infinite path $\mathfrak{r}$ 
that starts
at the root of ${\hat{\tt P}_{{\tt S(}\pi,{\tt \parystr)}}}$ and 
goes
through at least one configuration of the form $\langle S(\pi, {\tt \para \cons \parystr)}, R({\tt \para}) \rangle$, let $[S{\tt\bquote}]$ be the first occurrence of such a configuration in $\mathfrak{r}$. See \figurename~\ref{fig4:Prelim} below. Then the first pivot configuration after $[S{\tt\bquote}]$ in the continuation of the path $\mathfrak{r}$ is a configuration $[C]$ of one of the forms:
\begin{itemize}
 \item[(1)] {if the predicate $R(\xi_1)$ is 
satisfiable 
 then $[C]$ is of the form ${\tt L(} \pi_1, \parystr, \pi, \parystr{\tt )}$ and this occurrence of $[C]$ is covered by one of its ancestors;
            }
 \item[(2)] {if the predicate $R(\para) \wedge  (\para \neq  \xi_1)$ is 
satisfiable 
then $[C]$ is 
 an $[S{\tt \bquote}]$-child of the form ${\tt S(}\pi, \parystr{\tt )}$. 
 \hfill$\ednote$
            }
\end{itemize}
\end{lemma}

{
\begin{figure}[h]
\[
\def\objectstyle{\scriptstyle} 
\def\labelstyle{\scriptstyle}
\xymatrix @+2mm @R-25pt @C-0pt {
  *[o][F]{\mathbf{\bullet}} 
        \ar[r]^{} 
  & *+[F-:<3pt>]{\texttt{[S]}}
     \ar@{.>}[r]^{\mathfrak{r}}  
     &    \bullet
            \ar@{.>}[rrrr]^{}  
            \ar@{.>}[dr]^(.6){\mathfrak{r}} 
     &
     &
     &   
     &
     &          
\\
     &  
     &          
     &  *+[F-:<3pt>]{\texttt{[{\it S}\quote]}} \ar@{.}@/_0.2pc/[rr]
                \ar@{.>}[dr]^(.7){\mathfrak{r}} 
                \ar@{.}[ddddrr]
     &  
     & \ar@/^1.5pc/@{.}[dddd]
     &  \ar@/^0.5pc/@{.}[dd]
     & 
\\ 
     &  
     &          
     & 
     &  *+[F-:<3pt>]{\texttt{[{\it C}]}} 
                                   \ar@{.}@/_0.3pc/[urr]^(0.9){\texttt{\ \ }}
                                   \ar@{.>}[rr]^(.8){\mathfrak{r}}
                                   \ar@{.}@/^0.3pc/[drr]
     &   
     & 
     &          
\\ 
     &  
     & 
     &  
     &  \ar@/_0.7pc/@{.}[uu] 
     &
     &          
     &          
\\ 
     & & & & & & &
\\ 
     & & & & & & & 
\\ 
}
\]
\vspace{-20pt}
%
\caption{The 
             complete
             unfolding tree ${\hat{\tt P}_{{\tt S(}\pi,\ {\tt \parystr)}}}$.}\label{fig4:Prelim}
\end{figure}
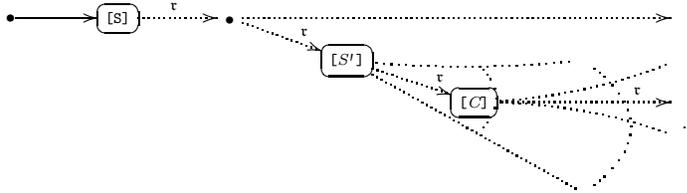
}

\section{Uniform Properties of the 
 Complete
Unfolding Tree of 
the Functional Program Model of the M-KMP Algorithm
}
In this section {\tt P} means the program considered in Introduction above.
Let $\mathfrak{t}$ be a path starting at the ${\hat{\tt P}_{{\tt S(}\pi, \parystr{\tt )}}}$-root, where $\pi\in{\cal A}^{+}$. 
Let the path $\mathfrak{t}$ correspond to a value of $\parystr$ of the form $\pi\zb\zb\concat\,\parxstr$.

The theorem below states that for any $\pi$ any nonempty prefix $\Psi$ of $\mathfrak{t}$ covers 
a subtree {\tt H} of  
${\hat{\tt P}_{{\tt S(}\pi,\ {\tt \parystr)}}}$ 
\st the {\tt H}-root is a leaf of the tree that 
 results 
from the last pivot configuration $[C] \in \Psi$ by means of 
the one-step unfolding 
 and does not belong to the path $\mathfrak{t}$. 
See Fig. \ref{fig5:UniformProp}. The corresponding parameter renamings, the covering morphisms, depend on concrete 
 configuration 
 pairs \--- the covering and covered ones.
Since the subtree {\tt H} corresponds to the first failure, when a symbol of the string $\parystr$ does not meet a symbol of the pattern $\pi$, then the 
edge {\hfill}incoming {\hfill}in {\hfill}the {\hfill}{\tt H}-root {\hfill}is {\hfill}labeled {\hfill}with {\hfill}a {\hfill}narrowing {\hfill}of {\hfill}the {\hfill}form\\ 
\zb$\parystr$\contr$\parb\,\parystr{\hspace{-4.3mm}}\quote$\verb|  |, this arrow should be read as {\it ``is of the form''},  
belonging to the path from the covering configuration to the covered one.
The covering pivot $L$-configurations from $\Psi$ include explicitly, \ie as constant terms, the main invariant of the M-KMP algorithm.

\subsection{The Main Contribution}
We have proved the following theorem. 
See also Fig. \ref{fig5:UniformProp}.

\begin{theorem}[On Covering]\label{th:on-covering}
Let a word $\pi\in{\cal A}^{+}$ and ${\hat{\tt P}_{{\tt S(}\pi, \parystr{\tt )}}}$ be given. Let $\mathfrak{t}$ be the first, the shortest,
 path from the ${\hat{\tt P}_{{\tt S(}\pi, \parystr{\tt )}}}$-root to a leaf labeled with the configuration \True, $\Psi$ \--- a nonempty prefix of the path $\mathfrak{t}$, $[C]$ \--- the last pivot configuration in $\Psi$. Then for any subtree {\tt H} of 
 ${\hat{\tt P}_{{\tt S(}\pi, \parystr{\tt )}}}$,
 rooted in a leaf of a tree resulted from
  $[C]$ 
  by means of 
  the one-step unfolding
  and not belonging to the path $\mathfrak{t}$, the prefix $\Psi$ covers {\tt H}.
\hfill$\ednote$
\end{theorem}

{
\begin{figure}[h]
\hspace{57pt}$\overbrace{\texttt{\hspace{190pt}}}^{\Psi}$
\vspace{-10pt}
\[
\def\objectstyle{\scriptstyle} 
\def\labelstyle{\scriptstyle}
\xymatrix @+2mm @R-25pt @C-0pt {
  *[o][F]{\mathbf{\bullet}} 
        \ar[r]^{} 
  & *+[F-:<3pt>]{\texttt{[S]}}
     \ar@{.>}[rr]^{\mathfrak{t}}  
     &    
     &    *+[F-:<3pt>]{\texttt{[{\it C}]}}
            \ar[r]^{\mathfrak{t}}  
            \ar[dr] 
     & \mathbf{\bullet}
            \ar@{.>}[rrr]^{\mathfrak{t}} 
     &   
     &          
     &  *+[o][F:green]{\texttt{\True}} 
\\
     &  
     &          
     & 
     &  *+[F-:<3pt>]{\texttt{[{\it C$_1$}]}} \ar@{.}[rr]
                \ar@{.>}[ddrr]^{\mathfrak{r}} 
                \ar@{.}[dddrr]
     &  
     & \ar@/^1.5pc/@{.}[ddd]
     &  
\\ 
     &  
     &          
     & 
     &          
     &  
     &  *+{{\tt H}} 
     &  
\\ 
     &  
     & 
     &  
     &  
     &
     &  
     &          
\\ 
     & & & & & & &
\\ 
     & & & & & & & 
\\ 
}
\]
%
\vspace{-40pt}
\caption{The 
             complete
             unfolding tree ${\hat{\tt P}_{{\tt S(}\pi,\ {\tt \parystr)}}}$.}\label{fig5:UniformProp}
\end{figure}
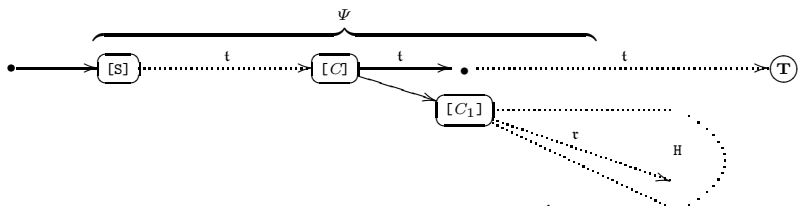
}

The On-Covering Theorem means that for any $\pi\in{\cal A}^{+}$ no generalization happens during supercompilation of the task of interest and there are finitely many the configurations in ${\hat{\tt P}_{{\tt S(}\pi, \parystr{\tt )}}}$ modulo 
 parameter
renaming.

Furthermore, 
the renaming folding substitutions include neither constant, static data, nor repeated parameter, 
dynamic variable, 
which are necessary in order to generate an accumulator in the residual program, keeping the passed track needed for backtracking along the input string. For an example see the repeating occurrence of the variable {\tt y} in the original program {\tt P}, the {\it rhs} of the first rule of the function {\tt S}. Since the residual function names are generated using the entire constant structure of the corresponding pivot configurations, removing the used structures,\footnote{For example the following pivot configuration \\ 
{\tt [L$_4$]:} {\tt L(\quote{bcaca}\quote,$\parystr$,\quote{abcabcaca}\quote,\quote{bca}\quote$\zb\zb\concat\,\parystr$)} 
will be transformed in the input format 
{\tt F{$\llcorner${\quote{bcaca}\quote,\quote{abcabcaca},\quote{bca}\quote$\zb\zb\concat\,repeated$-y$_{\tt str}$}$\lrcorner$}(y$_{\tt str}$)}
of the  residual function 
{\tt F{$\llcorner${\quote{bcaca}\quote,\quote{abcabcaca},\quote{bca}\quote$\zb\zb\concat\,repeated$-y$_{\tt str}$}}$\lrcorner$}, 
where 
the paired corner brackets $\llcorner{\tt arg}\lrcorner$ 
stand for encoding their {\tt arg} with a natural number.
}
the reader taking into account that the original program {\tt P} is a tail recursive may conclude that, actually, the {\it rhs} of any residual rewriting rule includes no constant data at all.

The above reasoning implies immediately that there is no backtracking in the residual program.
The function $f$ defined in \sectionname~\ref{Sec:M-KMP} is incorporated into the left-hand sides of the corresponding residual program running in 
${\cal O}(|\parystrzero|)$ time.

\vspace{5ex}
\paragraph{On computational complexity:}
While the worst-case time complexity of the original program {\tt P} is 
${\cal O}(|\parpatt| \times |\parystr|)$ the initial configuration $\Go(\pi_0,\parystr)$ of interest runs in linear time ${\cal O}( |\parystr|)$ for any $\pi_0 \in {\cal A}^{*}$. The corresponding residual program also runs in linear time. From a theoretical point of view, such a result is almost nothing. Nevertheless the result we have presented above shows that supercompilation using the Higman-Kruskal relation transforms the tail recursive program model {\tt P} to a specialized version of the M-KMP algorithm. The na{\ddi}ve algorithm differs meaningfully from the M-KMP algorithm and the last one is based on a quite nontrivial observation. See Sec. \ref{Sec:M-KMP}.   

The\hfill formal\hfill structures\hfill of\hfill the\hfill two\hfill algorithms\hfill differ\hfill as\hfill well.\hfill The\hfill first\hfill one\\ 
is\hfill tail\hfill recursive\hfill $\Go({\tt p},{\tt y})$\hfill while\hfill the\hfill second\hfill one\hfill is\hfill a\hfill composition\hfill of\hfill the\hfill form\\
 ${\tt M{\text -}KMP}(f({\tt p}),{\tt y})$ based on call-by-value evaluation. The supercompiler using the Higman-Kruskal relation is able to recognize that for any fixed pattern $\pi_0$ the backtracking loops in computing $\Go(\pi_0,{\tt y})$ terminate and therefore can be completely unfolded. That in turn allows the supercompiler to noticeably improve the constant factor in the upper bound on the number of the interpretation steps of the residual program looking for well structured patterns. For the set of such a kind of patterns the constant factor does matter both in the practice of programming and programming-language theory \cite{BenAmram-Jones:ConstDoesMatter,Jones:Complex}.

\paragraph{Future Work}
It will be interesting to automatically generate some other efficient algorithms from na{\ddi}ve program models solving the same tasks. For example, discovering periodicities in strings \cite{Slisenko:M-KMP-Eng}.

It would also be to interestingly investigate the average time complexity of the residual program of interest, which is more relevant to the practice as compared to the worst-case time complexity.

\section*{Acknowledgement}
 The author would like to thank the anonymous referees, 
 whose thoughtful comments helped to improve the presentation of these results.
 
  I also  would like to thank Antonina Nepeivoda for her critical reading of the paper.

\bibliographystyle{splncs04}{}

\end{document}